\documentclass[doublecol]{epl2} 

\title{Quantum state transfer by time reversal in the continuum}
\shorttitle{Adiabatic quantum state transfer ... } 

\author{S. Longhi \inst{1,2}}
\shortauthor{S. Longhi \etal}

\institute{                    
  \inst{1}  Dipartimento di Fisica, Politecnico di Milano, Piazza L. da Vinci 32, I-20133 Milano, Italy\\
  \inst{2}  Istituto di Fotonica e Nanotecnlogie del Consiglio Nazionale delle Ricerche, sezione di Milano, Piazza L. da Vinci 32, I-20133 Milano, Italy}

\abstract{A method for high-fidelity quantum state transfer in a quantum network coupled to a continuum, based on time reversal in the continuum after decay,
is theoretically suggested. Provided that the energy spectrum of the network is symmetric around a reference energy and symmetric energy states are coupled the same way to the common continuum, ideal perfect state transfer can be obtained after time reversal. In particular, it is shown that in a linear tight-binding chain a quantum state can be transformed into its mirror image with respect to the center of the chain after a controllable time. As compared to quantum mirror image based on coherent transport in a static chain with properly tailored inhomogeneous hopping rates, our method does not require hopping rate engineering and is less sensitive to disorder for long transfer times.}
\begin{document}

\maketitle

\section{Introduction}

Perfect quantum state transfer (PST) in one-dimensional lattices or spin chains is a fundamental 
task for future quantum computation and  communications systems \cite{ref1,ref2,ref3,ref4}. Over the last decade, 
different quantum state transfer schemes have been proposed \cite{ref3,ref5,ref5bis,ref6,ref7,ref8,ref8bis,ref8tris,ref9,ref9bis,ref10,ref11,ref12,ref13,ref14,ref15,ref16,ref17,ref18,ref19}, including probabilistic state transfer in
a chain with uniform parameters \cite{ref5,ref9bis}, PST in time-independent chains with properly-tailored hopping amplitudes \cite{ref5bis,ref6,ref7,ref8,ref8bis,ref8tris,ref9},
 and state transfer using externally applied time-dependent control fields \cite{ref10,ref11,ref12,ref13,ref14,ref15,ref16,ref17,ref18,ref19}.  
Static Hamiltonians with specially tailored hopping rates offer one of the most appealing way to realize PST and spatial mirror image \cite{ref6,ref7}, since they do not require external control fields and avoid interactions with the system except at initialization and read-out. Recently, mirror-image PST Hamiltonians of such kinds have been experimentally realized for photons using coupled optical waveguides with engineered coupling constants \cite{ref20,ref21}. The fidelity of the transfer, however, is 
very sensitive to imperfections of the underlying Hamiltonian and can be significantly degraded. Moreover a careful timing of the interaction is required, preventing the possibility to delay the transfer process on demand.\par
In this Letter it is suggested a rather general method of perfect state transfer in a quantum network with controllable delay by coupling the system to a continuum of states and time-reversing the dynamics in the continuum on demand. As an example, we apply the technique to realize mirror-image PST in a uniform tight-binding chain coupled to a semi-infinite quantum wire, which does not require any tailoring of the hopping amplitudes.\\  
Time reversal symmetry is a fundamental concept in science, which has been widely exploited in several areas of classical and quantum physics ranging from quantum chaos \cite{ref22,ref23}, matter waves \cite{ref22,ref24}, electromagnetic waves \cite{ref25,ref26,ref27}, nonlinear and quantum optics \cite{ref28,ref29,ref30},  hydrodynamic and acoustic waves \cite{ref31,ref32}, etc.  Time reversal symmetry can be used for combating decoherence \cite{ref33}, for dynamical decoupling \cite{ref34}, for cooling of matter waves \cite{ref35}, for the generation of rouge waves \cite{ref36}, for communication systems and adaptive optics \cite{ref27,ref28}, and for controlling spontaneous photon emission and absorption of a matter qubit \cite{ref30}. The main aim of this Letter is to show how time reversal symmetry can be fruitfully exploited for quantum state transfer. The basic idea can be described as follows. Let us put the quantum network S, prepared in some quantum state $|\psi(t=0) \rangle_S$ at initial time $t=0$, in interaction with a continuum of states, described by the Hamiltonian $\hat{H}_C$. After some time, the initial excitation of the system S is completely decayed into the continuum of states, i.e. fully decoherence is attained. At a subsequent time $t_0$, time reversal of the continuum is achieved by replacing $\hat{H}_C$ by $-\hat{H}_C$, while the interaction Hamiltonian $\hat{H}_I$ and the system Hamiltonian $\hat{H}_S$ {\it are not} time reversed. Under certain symmetry conditions on $\hat{H}_I$ and $\hat{H}_S$, at the time $t_f=2t_0$ the excitation is fully refocused into the system S, however the final state $ | \psi(t_f) \rangle_S$ generally differs from the original one. In other words, time reversing the continuum leads to re-excitation of the system, but in a different state: a quantum state transfer, via time reversal of a continuum, is thus realized.

\begin{figure}
\onefigure[width=8cm]{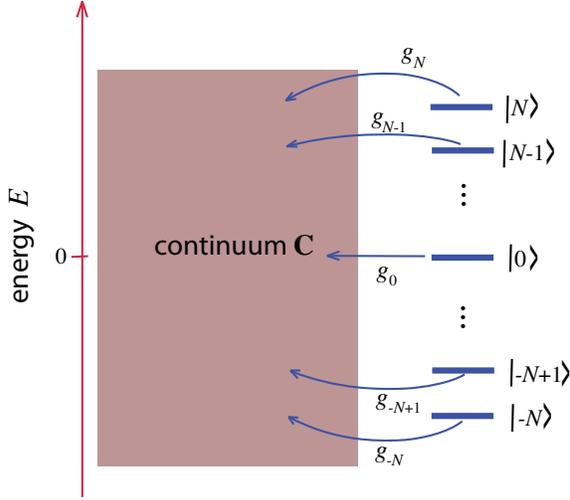}
\caption{(Color online) Schematic of $(2N+1)$ discrete states $|n \rangle$ coupled to a common continuum of states C.}
\end{figure}

\section{State transfer via time reversal of the continuum: general analysis} .\par
 Let us consider a quantum network S, which is composed by an odd number of sites $(2N+1)$ and described in the eigenstate basis $|n\rangle$ by the Hamiltonian (taking $\hbar=1$)
 \begin{equation}
 \hat{H}_S=\sum_{n=-N}^{N} \omega_n |n \rangle \langle n | 
 \end{equation}
 where $\omega_n$ is the eigenenergy of $|n \rangle$. The case of an even number of states is obtained by simply missing the $n=0$ state from the sum in Eq.(1). 
 In a typical PST setting, S describes a linear tight-binding chain, however the present analysis holds for a rather general quantum network with finite dimensionality. We assume that the network S is coupled to a continuum C of states described by the Hamiltonian
 \begin{equation}
 \hat{H}_C=\int d \alpha \;  \omega(\alpha) |\alpha \rangle \langle \alpha | 
 \end{equation}
   where $ \alpha$ is a continuous quantum number (or a continuous set of quantum numbers) and $\omega(\alpha)$ the energy dispersion relation of the continuum (scattered) states $| \alpha \rangle$. The interaction Hamiltonian is of the form
 \begin{equation}
 \hat{H}_I= \sum_{n=-N}^{N} \int d \alpha \left[ g_n(\alpha) |n  \rangle \langle \alpha |+ H.c. \right] 
 \end{equation}
  where $g_n(\alpha)$ are the coupling functions (see Fig.1). If we expand the quantum state of the full system $\{$S+C$\}$ as 
  \begin{equation}
  | \psi(t) \rangle= \sum_{n=-N}^N c_n(t) | n \rangle + \int d \alpha \Phi(\alpha,t) | \alpha \rangle,
  \end{equation}
  the evolution equations for the amplitude probabilities $c_n(t)$ and $\Phi(\alpha,t)$, as obtained from the Schr\"{o}dinger equation $ i \partial_t | \psi(t) \rangle= \hat{H} | \psi(t) \rangle$ with $\hat{H}= \hat{H}_S+\hat{H}_C+\hat{H}_I$, read
  \begin{eqnarray}
  i \dot{c}_n(t) & = & \omega_n c_n+\int d \alpha g_n(\alpha) \Phi(\alpha,t) \\
  i \dot{\Phi} (\alpha,t) & = & \omega(\alpha) \Phi(\alpha,t)+\sum_{n=-N}^{N} g^*_n(\alpha) c_n(t),
  \end{eqnarray} 
  where the dot indicates the derivative with respect to time $t$.
  To realize quantum state transfer, let us assume that:\\
  (i) After a suitable choice of the zero energy, the energy spectrum of S is symmetric at around $\omega=0$, i.e. 
  \begin{equation}
  \omega_{-n}= -\omega_n
  \end{equation} 
  ($n=0,1,2,...,N$). In particular, for an odd number of discrete states in S the above condition implies $\omega_0=0$.\\
   (ii) The following symmetry relations hold among the coupling functions
   \begin{equation}
   g_{-n}(\alpha)=g_n(\alpha),
   \end{equation}
   i.e. the states $| n \rangle$ and $|-n \rangle$ are coupled in the same way to the continuum.\\
   (iii) There are not bound states, neither inside nor outside the continuum (the condition for the absence of bound states in a multilevel system is discussed in \cite{ref37}). \\
 At initial time, let us prepare the system S in a certain quantum state, given by a coherent superposition of the eigenstates $|n \rangle$ with amplitudes $c_n(0)$; no excitation in the continuum is typically assumed,  i.e. $\Phi(\alpha,0)=0$, however such a constraint is not necessary.  Owing to the assumption (iii), the initial excitation of the system S fully decays into the continuum, i.e. for a sufficiently long time $t_0$ one has $c_n(t_0) \simeq 0$, with the entire initial excitation being transferred into the continuum ($\Phi(\alpha,t_0) \neq 0$). The time $t_0$ must be chosen larger than the characteristic decay time, which in the Weisskopf-Wigner approximation is basically determined by the coupling strength between S and C and by the density of states of C at the energies $\omega_n$; in the strong coupling regime transient back-flow into the system can arise before full decay, however such a regime can be suitably avoided. At the time $t=t_0$, we time-reverse the dynamics in the continuum by flipping the sign of $\hat{H}_C$, i.e. by the replacement $\omega(\alpha) \rightarrow - \omega(\alpha)$ in Eq.(6). However, the Hamiltonians $\hat{H}_S$ and $\hat{H}_I$ remain unchanged, i.e. they are not time reversed. Physical implementations of the time reversal will be discussed in the next section for a specific example of the continuum. Owing to the additional assumptions (i) and (ii), for $t>t_0$ it can be readily shown that the solution to Eqs.(5) and (6) is given by
 \begin{eqnarray}
 c_n(t) & = & -c_{-n}(2t_0-t) \\
 \Phi(\alpha,t) & = & \Phi(\alpha,2t_0-t)
 \end{eqnarray}
 which is an exact result provided that $c_{-n}(t_0)=-c_{n}(t_0)$, i.e. provided that $t_0$ is sufficiently larger than the decay time in the continuum so as $c_n(t_0) \simeq 0$. However, no upper limit to $t_0$ is in principle required. At the final time $t_f=2t_0$, from Eqs.(9) and (10) it follows that $c_n(t_f)=-c_{-n}(t_f)$  and $\Phi(\alpha,t_f)=\Phi(\alpha,0)$, i.e. the excitation has been fully refocused into the system S {\it but} in a mirror-symmetric quantum state owing to the reversal $n \rightarrow -n $ in Eq.(9). It should be pointed out that the method holds also when $H_C$ describes a discrete energy spectrum with many and closely-spaced energy levels (a good approximation of a continuum), however in such a case an upper limit to the transfer time $2t_0$ will arise according to the Poincare quantum recurrence theorem.
 
 \begin{figure}
\onefigure[width=8.5cm]{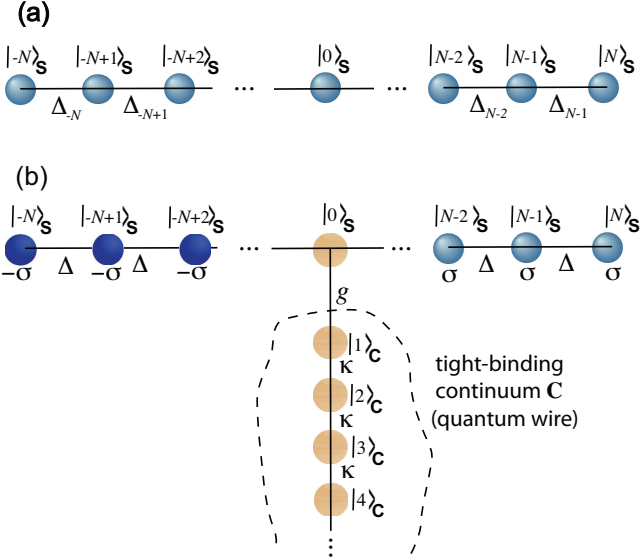}
\caption{(Color online) Mirror-image in a linear tight-binding quantum chain S comprising $(2N+1)$ sites. (a) Mirror-image based on a chain with inhomogeneous hopping rate distribution defined by Eq.(11) \cite{ref6,ref7,ref8tris}. (b) Mirror-image in a linear chain with homogeneous hopping rate based on coupling to a tight-binding continuum (semi-infinite quantum wire).}
\end{figure}
 
 \section{Spatial-mirror quantum image in a tight-binding chain} As an application of the time-reversal method described in the previous section, let us implement spatial mirror quantum image in a tight-binding linear chain S made of $(2N+1)$ sites $|n \rangle_S$, $|n| \leq N$. Mirror image in such a system can be realized by proper tailoring the hopping rates between adjacent sites in the chain, as suggested and demonstrated in several previous works \cite{ref6,ref7,ref8,ref8bis,ref8tris,ref20}; see Fig.2(a). Let us assume that the hopping rate $\Delta_n$ between sites $|n \rangle_S$ and $| n+1 \rangle_S$ is tailored such that
 \begin{equation}
 \Delta_n=\Delta \sqrt{(n+N+1)(N-n)}
 \end{equation}
 ($n=-N,-N+1,....,N-1$). Then, it can be shown \cite{ref6,ref7}  that the time $t_f$ for mirror image is given by an odd multiple than
\begin{equation}
\tau=\frac{\pi}{2 \Delta}.
\end{equation}
This means that, if at time $t=0$ one has $|\psi(t=0) \rangle_S=|l \rangle_S$, than $|\psi(t_f) \rangle_S=|-l \rangle_S$  (mirror image) whenever $t_f=(2s+1) \tau$ ($s$ integer), whereas $|\psi(t_f) \rangle_S=|l \rangle_S$ (self-image) whenever $t_f=2s \tau$.
 Here we follow a different route, which does not require inhomogeneous hopping rates and enables to delay the transfer time on demand. To this aim, we couple the chain S with a continuum of states consisting of a semi-infinite tight-binding quantum wire C with Wannier sites $|l \rangle_C$, $l=1,2,3,...$; see Fig.2(b). The chain S is attached to the quantum wire C via its central site $|0\rangle_S$ with a hopping amplitude $g$ to the site $| 1 \rangle_C$. The hopping rates in the chain S and in the quantum wire C are uniform and are denoted by $\Delta$ and $\kappa$, respectively. The semi-infinite quantum wire realizes a tight-binding continuum with energy $\omega(\alpha)=2 \kappa \cos (\alpha)$, where $\alpha$ is the Bloch wave number varying in the range $0 \leq \alpha \leq \pi$, and with Bloch states $| \alpha \rangle=(2 \pi)^{1/2}\sum_{l=1}^{\infty} \sin(l \alpha) |l \rangle_C$ (see, for instance, \cite{ref38}). The site energies $\sigma_n$ of the chain S are assumed to be antisymmetric with respect to the center of the chain $n=0$, i.e. $\sigma_{-n}=-\sigma_n$, with zero energy taken as a reference at the center of the tight-binding continuum band. Typically we will assume
 \begin{equation}
 \sigma_n= \left\{
 \begin{array}{cc}
 -\sigma & N \leq n<0 \\
 0 & n=0 \\
 \sigma & 0<n \leq N
 \end{array}
 \right.
 \end{equation}
 Under such an assumption, it readily follows that the energy spectrum $\omega_n$ of the chain S is symmetric with respect to $\omega=0$. In fact, if $|n \rangle=\sum_{l=-N}^{N} a_l^{(n)} |l \rangle_S$ is an eigenstate of S with energy $\omega_n$, then $|-n \rangle=\sum_{l=-N}^{N}(-1)^l a_{-l} |l \rangle_S$ is an eigenvector of S as well with energy $-\omega_n$. Moreover, since the two eigenvectors $| \pm n \rangle$ have the same occupation amplitude $a_0$ at the site $|0 \rangle_S$ coupled to the quantum wire C, it follows that they have the same coupling amplitude $g_n(\alpha)=g_{-n}(\alpha)$ with the Bloch modes $| \alpha \rangle$ of the quantum wire C. The conditions for the time reversal procedure described in the previous section are thus met. Parameter values should be chosen to avoid the existence of bound states, either outside or inside the continuum, when the chain S in connected to the quantum wire C. Bound states outside the continuum are avoided provided that $2 \kappa$ is sufficiently larger than $2\Delta+\sigma $, i.e. provided that the energies $\omega_n$ of S, bounded in the range $(-2\Delta-\sigma, 2\Delta + \sigma)$, are embedded into the continuous energy band $(-2 \kappa, 2 \kappa)$ of the quantum wire.  To avoid bound states inside the continuum, $\sigma / \kappa$ should not take the forbidden values $\cos[ l_1 \pi/(N+1)]-\cos[l_2 \pi / (N+1)]$ for $l_1,l_2=1,2,...,N$, corresponding to effective decoupling of S from the quantum wire C at the node $|0 \rangle_S$. To realize time reversal of the continuum C, it is enough to flip the sign of the hopping rate $\kappa \rightarrow -\kappa$ at time $t_0$. This can be accomplished using different methods, for example by fast half-cycle Bloch oscillations or by imprinting a $\pi$ phase shift to adjacent amplitudes, as suggested and demonstrated  for cold atoms in optical lattices and for photons in evanescently-coupled dielectric waveguide arrays in a few recent works \cite{ref24,ref39,ref40,ref41,ref42,ref43}. Interestingly, owing to the symmetry properties of the eigenstates $|n \rangle$ of S, if the system is initially prepared in a Wannier state, $| \psi(t=0) \rangle_S=|l\rangle_S$, then at the time $t_f=2 t_0$ one has $|\psi(t_f) \rangle_S=|-l \rangle_S$, i.e. mirror image has been realized.
  \begin{figure}
\onefigure[width=8.5cm]{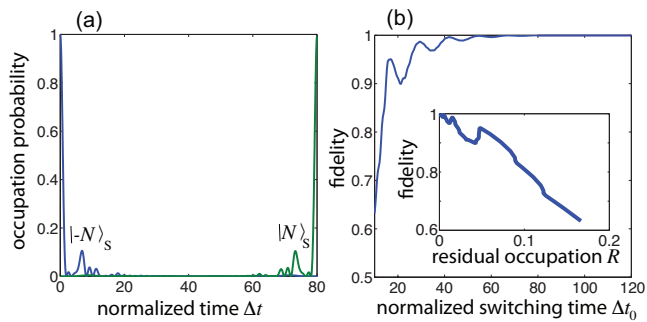}
\caption{(Color online) Example of quantum state transfer in a chain of $(2N+1)=11$ sites coupled to a quantum wire [Fig.2(b)]. (a) Temporal evolution of the occupation probabilities of the initially-excited state $|-N \rangle_S$ and of its mirror image state $|N \rangle_S$; time reversal in the continuum is set at time $t_0=40 / \Delta$. (b) Transfer probability to the mirror-image site $|N \rangle_S$ (fidelity) as a function of the reversal time $t_0$. Parameter values are: $\kappa / \Delta=2$, $g/ \Delta=2$, and $ \sigma / \Delta = 3/2$. The inset in (b) depicts the dependence of the fidelity on the residual occupation probability $R$ of the chain S at the switching time $t_0$.}
\end{figure}
  \begin{figure}
\onefigure[width=8.5cm]{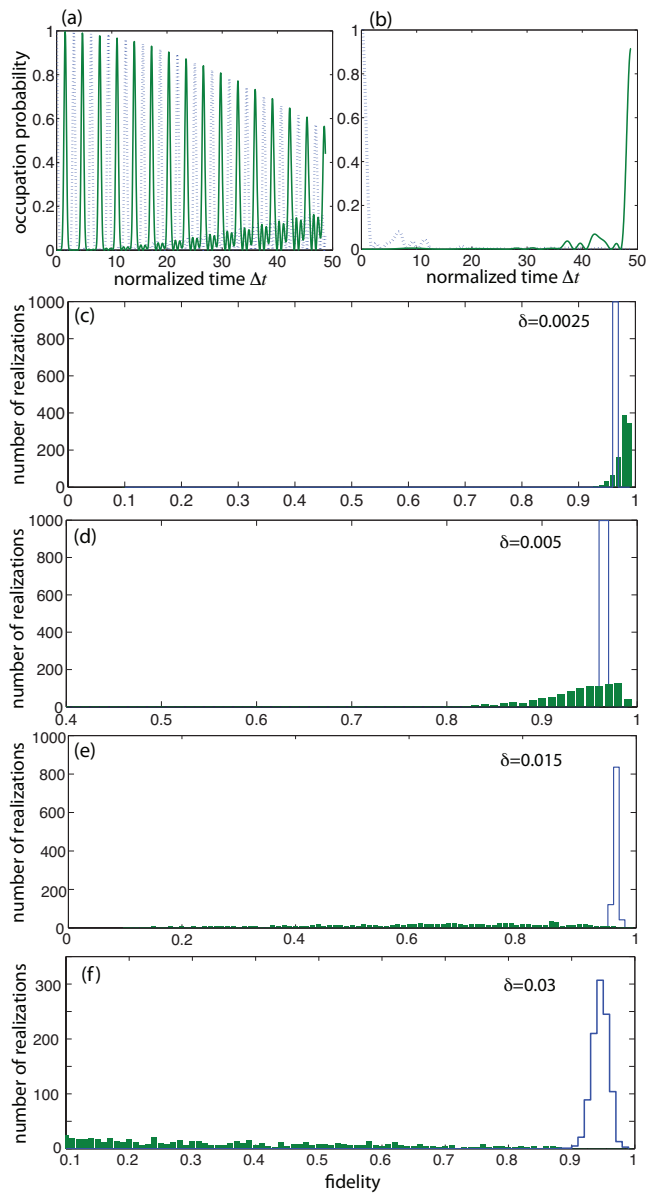}
\caption{(Color online) Effect of disorder of hopping rates on the quantum state transfer process in a liner chain made of $(2N+1)=11$ sites using the two methods shown in Fig.2(a) and 2(b). Parameter values are given in text. (a,b) Example of quantum state transfer (evolution of the occupation probabilities of initially excited site $|-N \rangle_S$, dotted curve, and of its mirror-image state $|N \rangle_S$, solid curve) for the schemes of Fig.2(a) and (b). Disorder strength is $\delta=0.03$, transfer time $t_f=31 \tau$. (c-f) Distributions of the transfer probability (fidelity) for 1000 realizations of disorder in the schemes of Fig.2(a) (bars) and Fig.2(b) (solid curve) for increasing values of disorder $\delta$.}
\end{figure}
 An example of the mirror image transfer for a chain S made of $(2N+1)=11$ sites is shown in Fig.3 for parameter values $\kappa / \Delta= g /\Delta=2$ and $\sigma / \Delta=3/2$. At initial time $t=0$ the chain S is excited in its left-edge state $|-N \rangle_S$, time-reversal of the continuum is realized at time $t_0$, and mirror image transfer is observed at $t_f=2 t_0$. Figure 3(a) shows the numerically-computed behavior of the occupation probabilities of site $|-N \rangle_S$ and of its mirror image $|N \rangle_S$ versus time for $t_0=40/\Delta$. As it can be seen, in the first stage of the dynamics ($0<t<t_0$) there is an almost complete decay of the excitation into the continuum; after time reversal of the continuum Hamiltonian at $t=t_0$, the excitation is almost completely refocused into the mirror-image site $|N\rangle_S$ of the chain S, with a fidelity of $\simeq 99.2 \%$. The fidelity of the transfer process typically increases by increasing the delay time $t_0$, reaching the value of almost $100\%$ for switching times $t_0$ larger than $ \sim 40/\Delta$; see Fig.3(b). On the other hand, the fidelity greatly degrades for short switching times $t_0$, since at such short times the initial excitation in the chain has not be fully decayed into the continuum. The inset is Fig.3(b) shows the behavior of the fidelity versus the residual occupation probability $R$ in the system S at reversal time $t_0$, defined as $R=\sum_{n=-N}^{N} |\langle \psi(t_0) | n \rangle_S |^2$. As expected, the fidelity degrades when $R$ is nonvanishing, i.e. for too short switching times $t_0$ at which residual excitation in the system is still present. However, there is not an upper limit to the switching time $t_0$, making the method appealing for storing the initial excitation in the continuum and release it on demand into the mirror-image site by time reversal of the continuum. \\
   \begin{figure}
\onefigure[width=8.5cm]{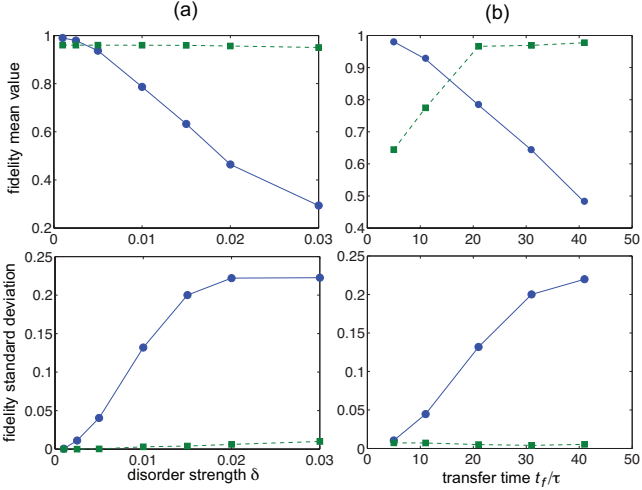}
\caption{(Color online) (a) Numerically-computed mean values (upper plots) and standard deviations (lower plots) of the fidelity versus disorder strength $\delta$ for a fixed transfer time $t_f=31 \tau$. (b)  Mean values (upper plots) and standard deviations (lower plots) of the fidelity versus transfer time $t_f$ for a fixed value of disorder $\delta=0.015$.  Circles and squares refer to the transfer schemes of Fig.2(a) and 2(b), respectively.}
\end{figure}
 It should be noted that the same goal of storing the excitation in the chain, even for long times, can be in principle  realized in the traditional scheme of Fig.2(a). To store the excitation for a time larger than the mirror-image transfer time $\tau$ defined by Eq.(12), one can just consider a final time $t_f$ of the dynamics to be an odd multiple than $\tau$. In this case the excitation oscillates between sites $|-N \rangle_S$ and $| N \rangle_S$ before being extracted from the bus chain at the final time $t_f$.  As compared to PST using the chain of Fig.2(a), the method of Fig.2(b) turns out to be less sensitive to imperfections and disorder for a long storage time $t_f$. We checked such a property by numerically computing the transfer probability in the two systems of Fig.2(a) and (b) in the presence of disorder in the hopping rates. The transfer time $t_f$ in the two cases is chosen to be the same and equal to an odd multiple than $\tau$, defined by Eq.(12). In this way, in the absence of disorder both systems store and release the excitation in the mirror-image state with high fidelity after the same time interval $t_f$. While in the system of Fig.2(a) there is an oscillatory behavior of excitation between the two mirror-image sites, in the system of Fig.2(b) the excitation is stored in the continuum and then refocused into the mirror-image site of the chain. Disorder of the hopping rates in the chain of Fig.2(a) is mimicked by multiplying the ideal hopping rate distribution  $\Delta_n$ [Eq.(11)] by a factor $f_n=1+\delta_n$, i.e. $\Delta_n \rightarrow \Delta_n f_n$, where $\delta_n$ is a random number with uniform distribution in the interval $(-\delta,\delta)$, with $\delta \ll 1$. In the case of Fig.2(b), disorder of hopping rates is assumed {\it both} in the main chain S and in the semi-infinite quantum wire C, multiplying the ideal (uniform) hopping rate distributions by the random factor $f_n=1+\delta_n$ with the same variance $\delta$ as in the previous case. Figures 4(a) and (b) show, as an example, the behavior of the occupation probabilities at sites $| \pm N \rangle_S$ in the two different transfer schemes for a given realization of disorder in the hopping rates  and for parameter values $(2N+1)=11$, $\kappa / \Delta= g /\Delta=2$, $\sigma / \Delta=3/2$, $t_f= 31 \pi/(2 \Delta)= 31 \tau$; disorder strength is $\delta =0.03$. Note that for the scheme of Fig.2(a) a number $t_f/\tau=31$ of oscillations are observed before the final extraction time $t_f$, and that the disorder in the hopping rates cumulatively degrades the perfect transfer behavior of the ideal system from one oscillation to the next one  [Fig.4(a)]. Such a cumulative effect greatly degrades the fidelity of the transfer process for the scheme of Fig.2(a), while for the scheme of Fig.2(b) the fidelity remains at satisfactorily levels in spite of the same degree of disorder. Such a property mainly stems from the fact that disorder of hopping rates in the semi-infinite quantum wire is basically cancelled by time reversal \cite{refReferee}, and the main impact of disorder on the fidelity results from disorder-induced symmetry breaking [Eqs.(7) and (8) are not rigorously satisfied] and possible back-scattering from the continuum into the main chain. To provide a comparative analysis of the impact of disorder on the fidelity of the transfer process, we performed an extended statistical analysis by computing the distribution of the transfer probabilities corresponding to 1000 different realizations of disorder for the two transfer schemes of Fig.2(a) and (b) and for a transfer time $t_f=31 \tau$. Figures 4(c-f) show a few examples of the probability distributions obtained from the statistical analysis as the strength of disorder increases. The figures clearly show that, while for small disorder the two schemes provide a high fidelity for the transfer process [Fig.4(c)], as the disorder strength $\delta$ is increased [Figs.4(d-f)]  the  fidelity of the transfer process for the scheme of Fig.2(b) remains at a satisfactorily level while the fidelity of transfer is fully degraded for the scheme of Fig.2(a). Figure 5 shows the numerically-computed behavior of the mean values and standard deviations of the probability distributions of the fidelity in the two different schemes for increasing disorder strength $\delta$ at a given transfer time $t_f=31 \tau$ [Fig.5(a)], and  for increasing transfer time $t_f$ at a given disorder strength $\delta=0.015$. Clearly, for long transfer times the scheme of Fig.2(b) outperforms the one of Fig.2(a). Interestingly, the fidelity for the scheme of Fig.2(b) remains high and  almost constant in the considered range of disorder strength, as shown in Fig.5(a);  to observe degradation of fidelity to e.g. $\sim 0.83$ a disorder strength $\delta \sim 0.1$ is needed.
\section{Conclusion} In this Letter  we have suggested the possibility to exploit time-reversal symmetry to realize perfect quantum state transfer in a quantum network coupled to a continuum of states. The idea has been applied to the design of a mirror-image quantum state transfer protocol in a linear chain coupled to a semi-infinite tight-binding quantum wire. The fidelity of the transfer process via time-reversal of the continuum has been compared to the one obtained in mirror-image quantum state transfer protocol of Refs.\cite{ref6,ref7}, based on a tight-binding chain with engineered hopping rates. The numerical results indicate that our method is less sensitive to disorder in hopping rates than the latter one when long storage times in the chain are considered. The proposed transfer scheme [Fig.2(b)], with time reversal of the semi-infinite quantum wire, can be implemented in evanescently-coupled optical waveguides with current available technology \cite{ref41}. It is envisaged that the idea of time reversal symmetry could be helpful in the design of other quantum state transfer protocols with high fidelity and with a controlled transfer time.



\end{document}